\documentclass{elsart}
\usepackage{graphicx}
\usepackage{amsmath}
\usepackage{amssymb}
\usepackage{epsfig}
\usepackage{dcolumn}
\usepackage{bm}
\begin{document}
\begin{frontmatter}

\title{Multifractal theory within quantum calculus}
\author{ Alexander Olemskoi}
\address{Institute of Applied Physics, Nat. Acad. Sci. of Ukraine \break 58, Petropavlovskaya St., 40030 Sumy,
Ukraine}
\author{Irina Shuda}
\address{Sumy State University \break 2, Rimskii-Korsakov St., 40007 Sumy, Ukraine}

\date{}

\begin{abstract}
Within framework of the quantum calculus, we represent the partition function
and the mass exponent of a multifractal, as well as the average of random
variables distributed over self-similar set, on the basis of the deformed
expansion in powers of the difference $q-1$. For the partition function, such
expansion is shown to be determined by binomial-type combinations of the
Tsallis entropies related to manifold deformations, while the mass exponent
expansion generalizes known relation $\tau_q=D_q(q-1)$. We find the physical
average related to the escort probability in terms of the deformed expansion as
well. It is demonstrated the mass exponent can acquire a singularity that
relates to a phase transition of the multifractal set in the course of its
deformation.
\end{abstract}

\end{frontmatter}
\maketitle

\section{Introduction}\label{Sec.1}

Fractal conception \cite{Mandel} has become widespread idea of contemporary
science (see Refs. \cite{Feder} -- \cite{Sor}, for review). Characteristic
feature of fractal sets is known to be the self-similarity: if one takes a part
of the whole set, it looks like the original set after a deformation
$\lambda\ne 1$. Formal basis of the self-similarity is the power-law function
$F\sim\lambda^{h}$ with a Hurst exponent $h$ (for time series, value $F$ is
reduced to the fluctuation amplitude and $\lambda$ is the interval size within
which this amplitude is determined). While the simple case of monofractal is
characterized by a single exponent $h$, a multifractal system is described by a
continuous spectrum of exponents (singularity spectrum) $h(q)$ \cite{multi}.
Along this way, a self-similarity degree $q$ represents the exponent of a
homogeneous function being characteristic function of self-similar systems
\cite{Erzan} (so, within nonextensive thermostatistics, this exponent expresses
the escort probability $P_i\propto p_i^q$ throughout the original one $p_i$
\cite{T,BS}). In physical applications, a central role is played by the
partition function $Z_q\propto l^{\tau(q)}$ ($l$ is a characteristic size of
boxes covering multifractal) to be determined by the mass exponent $\tau(q)$
connected with the generalized Hurst exponent $h(q)$ by the relation
$\tau(q)=qh(q)-1$.

Since fractals are scale invariant sets, it is natural to apply the quantum
calculus to description of multifractals because quantum analysis is based on
the Jackson derivative which yields variation of a function with respect to the
deformation of its argument \cite{QC,QG}. This Letter is devoted to such a
description with using deformed series of expansions of the partition function,
the mass exponent and the averages of random variables over deformed powers of
the difference $q-1$. As shows the consideration in Section \ref{Sec.2}, the
coefficients of the partition function expansion are reduced to binomial-type
combinations of the Tsallis entropies related to manifold deformations
$\lambda^m$ with growing powers $m=0,1,\dots$ Respectively, Section \ref{Sec.3}
shows that escort probability is generated by the mass exponent whose expansion
into deformed power series represents the generalization of well-known relation
$\tau_q=D_q(q-1)$ with fractal dimension spectra $D_q$ corresponding to
manifold deformations. According to Section \ref{Sec.4}, above expansion
permits also to expresses physical averages over the escort probability $P_i$
in terms of related averages based on the ordinary one $p_i$. Section
\ref{Sec.5} contains discussion of results obtained, and in Appendix we adduce
necessary information from the quantum calculus.

\section{Partition function}\label{Sec.2}

Following the standard scheme \cite{multi,Feder}, we consider a multifractal
set covered by boxes $i=1,2,\dots,W$ with number $W\to\infty$. The peculiarity
of self-similar sets is that the probability to occupy a box $i$ is determined
by its size $l_i\to 0$ according to the power-law relation $p_i=l_i^\alpha$
characterized by a H\"older exponent $\alpha>0$. All properties of a
multifractal is known to be determined by the partition function
\begin{equation}
Z_q=Z_q\{p_i\}:=\sum_{i=1}^W p_i^q
 \label{Z}
\end{equation}
which according to the normalization condition takes the value $Z_q=1$ at
$q=1$. Because $p_i\leq 1$ for all boxes $i$, the function (\ref{Z}) decreases
monotonically from maximum magnitude $Z_q=W$ related to $q=0$. In the simplest
case of the flat distribution $p_i=1/W$ fixed by the statistical weight $W\gg
1$, one has the exponential decay $Z_q=W^{1-q}$.

Our approach is based on the expansion of the partition function (\ref{Z}) into
the deformed series (\ref{7}) in powers of the difference $q-1$ \cite{QC}
\begin{equation}
Z_q^\lambda= -\sum\limits_{n=0}^\infty\frac{\mathcal{S}_{\lambda}^{(n)}}
{[n]_\lambda!}(q-1)_\lambda^{(n)}.
 \label{Z1}
\end{equation}
Here, the deformed powers $(q-1)_\lambda^{(n)}$ are determined by the binomial
(\ref{8}), while the kernels
$\mathcal{S}_{\lambda}^{(n)}=\mathcal{S}_{\lambda}^{(n)}\{p_{i}\}$ are defined
by the $n$-fold action of the Jackson derivative:
\begin{equation}
\mathcal{S}_{\lambda}^{(n)}=-\left.\big(q\mathcal{D}_q^\lambda\big)^n
Z_q\right|_{q=1}.
 \label{Z2}
\end{equation}
By using the method of induction with accounting the definitions (\ref{1}) and
(\ref{Z}), one obtains the explicit expression
\begin{equation}
\mathcal{S}_{\lambda}^{(n)}=
-(\lambda-1)^{-n}\sum\limits_{m=0}^n(-1)^{n-m}{n\choose m}Z_{\lambda^m}
 \label{kernel}
\end{equation}
where ${n\choose m}=\frac{n!}{m!(n-m)!}$ are binomial coefficients. In absence
of deformation $(\lambda=1)$, all coefficients (\ref{kernel}) equal zero, apart
from the single term $\mathcal{S}_1^{(0)}=-1$. Inserting these coefficients
into the series (\ref{Z1}) gives $Z_1=1$ as desired.

It is easily to see that the set of kernels (\ref{kernel}) with $n>0$ is
expressed in terms of the Tsallis entropy \cite{T}
\begin{equation}
S_\lambda\{p_i\}:=-\frac{\sum_i p_i^\lambda -1}{\lambda-1}=-\frac{Z_\lambda
-1}{\lambda-1}.
 \label{S}
\end{equation}
With growth of the deformation parameter $\lambda$, this entropy decreases
monotonically with inflection in the point $\lambda=1$ where the expression
(\ref{S}) takes the Boltzmann-Gibbs form $S_1=-\sum_i p_i\ln(p_i)$. In the case
$n=0$, the coefficient (\ref{kernel}) does not depend on the deformation
parameter $\lambda$ to give the value $\mathcal{S}^{(0)}_\lambda=-1$. At $n=1$,
one obtains \cite{Abe}
\begin{equation}
\mathcal{S}_\lambda^{(1)}\{p_i\}:=-\left.\frac{Z_{\lambda
q}-Z_q}{\lambda-1}\right|_{q=1}=-\frac{\sum_i p_i^\lambda
-1}{\lambda-1}=S_\lambda\{p_i\}.
 \label{E2}
\end{equation}
Respectively, in the second order $n=2$, Eq.(\ref{kernel}) yields
\begin{equation}
\mathcal{S}_{\lambda}^{(2)}\{p_i\}=-\frac{1-2\sum_i p_i^\lambda +\sum_i
p_i^{\lambda^2}}{(\lambda-1)^2}.
 \label{E4}
\end{equation}
With accounting the definition (\ref{S}) one has
\begin{equation}
\mathcal{S}_{\lambda}^{(2)}=-\frac{2}{\lambda-1}
S_{\lambda}+\frac{\lambda+1}{\lambda-1}S_{\lambda^2}.
 \label{E5}
\end{equation}
In arbitrary order $n$, combination of equations (\ref{kernel}) and (\ref{S})
arrives at the sum
\begin{equation}
\mathcal{S}_{\lambda}^{(n)}=\sum\limits_{m=0}^n(-1)^{n-m}{n\choose
m}\frac{\lambda^m-1}{(\lambda-1)^n}S_{\lambda^m}
 \label{K}
\end{equation}
which expresses in explicit form a generalized entropy of a multifractal by
means of contributions of the Tsallis entropies related to different powers of
the set deformation. In the limit $\lambda\to 0$, when $S_\lambda\sim W$,
accounting of all terms in the sum over $m=0,1,\dots,n$ arrives at cancellation
of the denominator $(\lambda-1)^n$ so that $\mathcal{S}_{0}^{(n)}\to W$;
similarly, at $\lambda\to 1$, one obtains the Boltzmann-Gibbs limit
$\mathcal{S}_{1}^{(n)}\to S_1=-\sum_i p_i\ln(p_i)$ as desired. Finally, in the
case $\lambda\to\infty$, where $S_{\lambda^m}\sim\lambda^{-m}$, the
term related to $m=0$ is reduced to zero so that main contribution gives the
term with $m=1$ to arrive at the sign-changing asymptotics
$\mathcal{S}_{\lambda}^{(n)}\sim(-1)^{n+1}\lambda^{-n}$.

For the flat distribution, when $Z_\lambda=W^{1-\lambda}$, the dependence
(\ref{S}) is characterized by the asymptotics $S_\lambda\sim
W\left[1-\ln(W)\lambda\right]$ in the limit $\lambda\ll 1$ and $S_\lambda\sim
1/\lambda$ at $\lambda\gg 1$. As a result, with the $\lambda$ growth the
coefficients (\ref{kernel}) increase from the value $\mathcal{S}_{0}^{(n)}=W$
to the Boltzmann magnitude $\mathcal{S}_1^{(n)}=W\ln(W)$ and then decay to the
asymptotics $\mathcal{S}_{\lambda}^{(n)}\sim(-1)^{n+1}\lambda^{-n}$.

\section{Characteristic function of escort probability}\label{Sec.3}

Following pseudo-thermodynamic picture of multifractal sets \cite{BS}, let us
define effective values of the free energy $\tau_q$, the internal energy
$\alpha$ and the entropy $f$:
\begin{equation}\label{fa}
\tau_q:=\frac{\ln(Z_q)}{\ln(l)},\quad\alpha:=-\frac{\sum_i P_i\ln
p_i}{\ln(1/l)},\quad f:=-\frac{\sum_i P_i\ln P_i}{\ln(1/l)}.
\end{equation}
Here, $l\ll 1$ is the size of characteristic box in phase space, $p_i$ and
$P_i$ are original and escort probabilities to be connected with the definition
\begin{equation}\label{prob}
P_i:=\frac{p_i^q}{\sum_i p_i^q}=\frac{p_i^q}{Z_q}.
\end{equation}
Inserting this equation into the second expression (\ref{fa}), one obtains the
well-known Legendre transformation \cite{multi,Feder}
\begin{equation}\label{L}
\tau_q=q\alpha_q-f(\alpha_q)
\end{equation}
where $q$ plays the role of the inverse temperature and the internal energy is
specified with the state equation
\begin{equation}\label{E}
\alpha_q=\frac{{\rm d}\tau_q}{{\rm d}q}.
\end{equation}

It is easily to convince the escort probability (\ref{prob}) is generated by
the characteristic function being the mass exponent
\begin{equation}
\tau_q:=\frac{\ln(Z_q)}{\ln(l)}=\frac{\ln\left(\sum_i p_i^q\right)}{\ln(l)}.
 \label{tau}
\end{equation}
Really, one has
\begin{equation}\label{P}
P_i=\frac{\ln(l)}{q}~p_i\frac{\partial\tau_q}{\partial
p_i}=q^{-1}\frac{\partial\ln(Z_q)}{\partial \ln(p_i)}.
\end{equation}
By analogy with Eq.(\ref{Z1}), one can expand the function (\ref{tau}) into the
deformed series
\begin{equation}
\tau_q^\lambda=
\sum\limits_{n=1}^\infty\frac{D_{\lambda}^{(n)}}{[n]_\lambda!}(q-1)_\lambda^{(n)}
 \label{tau1}
\end{equation}
being the generalization of the known relation $\tau_q=D_q(q-1)$ connecting the
mass exponent $\tau_q$ with the multifractal dimension spectrum $D_q$
\cite{Feder}. Similarly to the equations (\ref{Z2}) and (\ref{kernel}), the
kernel $D_{\lambda}^{(n)}$ is determined by the $n$-fold action of the Jackson
derivative:
\begin{equation}
D_\lambda^{(n)}:=\left.\big(q\mathcal{D}_q^\lambda\big)^n
\tau_q\right|_{q=1}=(\lambda-1)^{-n}\sum\limits_{m=0}^n(-1)^{n-m}{n\choose
m}\tau_{\lambda^m}.
 \label{DD}
\end{equation}
Noteworthy, the term with $n=0$ is absent in the series (\ref{tau1}). At $n=1$,
one reproduces the ordinary relation
$D_\lambda^{(1)}=\tau_{\lambda}/(\lambda-1)$, while the kernels
$D_{\lambda}^{(n)}$ with $n\geq 2$ include terms being proportional to
$\tau_{\lambda^m}/(\lambda-1)^n$ to correspond to manifold deformations
$\lambda^m$, $1<m\leq n$. Generally, the definition (\ref{DD}) yields a
hierarchy of the multifractal dimension spectra related to different
multiplication factors $n$ of the set deformation.

In trivial case of the flat distribution\footnote{It is worthwhile to stress
the expression (\ref{P}) is not applicable in this case because fixation of the
statistical weight $W$ makes impossible variation of the probability
$p_i=1/W$.}, when the partition function is $Z_\lambda=W^{1-\lambda}$, the
definition (\ref{tau}) yields the mass exponent $\tau_\lambda=D(1-\lambda)$
with the fractal dimension $D=\frac{\ln(W)}{\ln(1/l)}$ which tends to $D=1$
when the size of covering boxes $l$ goes to the inverse statistical weight
$1/W$. Thus, we can conclude the flat distribution relates to a monofractal
with dimension $D_\lambda=\frac{\ln(\lambda)}{\ln(1/l)}$ tending to a smooth
one-dimensional set at deformation $\lambda\to 1/l$.

\section{Random variable distributed over multifractal}\label{Sec.4}

Let us consider an observable $\phi_i$ distributed over a multifracal set.
According to Ref. \cite{T}, its mean value
\begin{equation}\label{O}
\left<\phi\right>:=\sum_i \phi_i P_i
\end{equation}
is determined by the escort probability (\ref{prob}). With accounting Eqs.
(\ref{P}) -- (\ref{DD}), the average (\ref{O}) can be expressed in terms of the
sum
\begin{equation} \label{OO}
\left<\phi\right>=\sum\limits_{n=0}^\infty
\frac{1}{[n]_\lambda!}\frac{(q-1)_\lambda^{(n)}}{(\lambda-1)^{n}}
\sum\limits_{m=0}^n(-1)^{n-m}{n\choose
m}\lambda^m\left<\phi\right>_{\lambda^m}
\end{equation}
where the specific average
\begin{equation} \label{A}
\left<\phi\right>_{\lambda^m}=\sum_i\phi_i P_i\left(\lambda^m\right)
\end{equation}
is introduced. At $m=0$, the escort distribution $P_i\left(\lambda^m\right)$ is
reduced to the original one $p_i$ so that related terms in the sum (\ref{OO})
are proportional to the ordinary average $\left<\phi\right>_1=\sum_i\phi_i
p_i$. On the other hand, in absence of deformation $(\lambda=1)$, the only term
with $n=0$ determines the mean value (\ref{O}):
$\left<\phi\right>=\left<\phi\right>_1$. In the limit $\lambda\to 0$, the
escort probability reduced to flat distribution $P_i(0)=1/W$ for arbitrary form
of the ordinary distribution so that average (\ref{O}) is determined by the
simple expression $\left<\phi\right>=W^{-1}\sum_i \phi_i$. Finally, in the case
$\lambda\to\infty$, main contribution is given by the terms with $m=0$ and
$m=1$ so that the average (\ref{OO}) takes the form
\begin{equation} \label{OO1}
\left<\phi\right>\simeq\left<\phi\right>_1+(q-1)
\big(\left<\phi\right>_\lambda-\lambda^{-1}\left<\phi\right>_1\big).
\end{equation}

\section{Conclusions}\label{Sec.5}

A principle peculiarity of above consideration is that the expansions
(\ref{Z1}), (\ref{tau1}) and (\ref{OO}) depend of both deformation parameter
$\lambda$ and self-similarity degree $q$. As shows the example of using the
flat distribution in the end of Section \ref{Sec.3}, the former determines the
fractal dimension $D_\lambda=\frac{\ln(\lambda)}{\ln(1/l)}$ in the simple case
of the monofractal generated by deformation $\lambda\gg 1$ of initial cell of
size $l\ll 1$. With passage to general case of multifractal, we obtain the
fractal dimension spectrum defined by the series (\ref{DD}). Along this line,
the self-similarity degree $q$ plays the role of a free parameter whose
variation describes the multifractal spectrum, while value of the deformation
parameter $\lambda$ fixes the scale of the multifractal resolution to be
determined by external conditions in the case of natural objects.

As pointed out in Introduction, our study is based on the Jackson derivative
(\ref{1}) which gives variation of a function with respect to scale choice of
its argument. This derivative expresses the coefficients of the deformed Taylor
expansion (\ref{7}) in powers of the deformed difference $q-1$ defined by the
binomial (\ref{8}). In this way, the expansion (\ref{Z1}) of the partition
function (\ref{Z}) is determined by the coefficients (\ref{K}) being
binomial-type combinations of the Tsallis entropies (\ref{S}) related to
manifold deformations $\lambda^m$ with growing powers $m=0,1,\dots$ According
to definitions (\ref{fa}) and (\ref{prob}) the mass exponent (\ref{tau}) is
expressed by the Legendre transformation (\ref{L}) to generate the escort
probability (\ref{P}). On the other hand, the expansion of the mass exponent
into deformed power series (\ref{tau1}) represents generalization of known
relation $\tau_q=D_q(q-1)$ with fractal dimension spectra (\ref{DD})
corresponding to manifold deformations. Moreover, making use of the deformed
expansion (\ref{7}) arrives at the expression (\ref{OO}) of the physical
averages (\ref{O}) corresponding to the escort probability $P_i$ in terms of
the averages (\ref{A}) related to multiple deformations.

Under deformation $\lambda$, above multifractal characteristics vary in the
following manner. The generalized entropies (\ref{K}) tend to the statistical
weight $W$ in the limit $\lambda\to 0$, take the Boltzmann-Gibbs magnitude
$\mathcal{S}_{1}^{(n)}=-\sum_i p_i\ln(p_i)$ at $\lambda=1$ and arrive at the
sign-changing asymptotics
$\mathcal{S}_{\lambda}^{(n)}\sim(-1)^{n+1}\lambda^{-n}$ when
$\lambda\to\infty$. What about the relation (\ref{OO}) between the physical
average $\left<\phi\right>=\sum_i \phi_i P_i$ and the ordinary one
$\left<\phi\right>_1=\sum_i \phi_i p_i$, one has the simple expression
$\left<\phi\right>=W^{-1}\sum_i \phi_i$ in the limit $\lambda\to 0$, the
trivial relation $\left<\phi\right>=\left<\phi\right>_1$ in absence of
deformation $(\lambda=1)$, and the difference
$\left<\phi\right>-\left<\phi\right>_1$ being proportional to the factor $q-1$
at $\lambda\to\infty$.

From physical point of view, the most tractable is the expansion of the mass
exponent (\ref{tau1}) whose coefficients represent fractal dimension spectra.
As shows the equality (\ref{DD}), the fractal dimensions $D^{(n)}_\lambda$
related to different powers $n$ of the series (\ref{tau1}) can change their
signs with variation of the deformation $\lambda$. Because, within the
pseudo-thermodynamic formalism, the mass exponent $\tau$ and the
self-similarity degree $q$ relate to the free energy and the inverse
temperature respectively, such a change means that thermodynamic potential can
obtain a singularity at some deformations $\lambda$. Within framework of the
physical presentation \cite{BS}, this singularity means the phase transition in
the course of the deformation of the multifractal set.

\section*{Appendix}\label{Sec.6}

Quantum analysis is known to be based on the Jackson derivative \cite{QC,QG}
\begin{equation}
\mathcal{D}_x^\lambda:=\frac{\lambda^{x\partial_x}-1}{(\lambda-1)x},\quad
\partial_x\equiv\frac{\partial}{\partial x}
 \label{1}
\end{equation}
where $\lambda$ is a deformation parameter. The deformed Taylor expansion
reads:
\begin{equation} \label{7}
\begin{split}
f(x)=f(a)+\frac{(x-a)_\lambda^{(1)}}{[1]_\lambda!}\left.\mathcal{D}_x^\lambda
f(x)\right|_{x=a}+\frac{(x-a)_\lambda^{(2)}}{[2]_\lambda!}\left.\big(\mathcal{D}_x^\lambda\big)^2
f(x)\right|_{x=a}+\dots\\=\sum\limits_{n=0}^\infty\frac{(x-a)_\lambda^{(n)}}{[n]_\lambda!}
\left.\big(\mathcal{D}_x^\lambda\big)^n f(x)\right|_{x=a}.
\end{split}
\end{equation}
Here, the deformed binomial
\begin{equation} \label{8}
\begin{split}
(x+y)_\lambda^{(n)}=(x+y)(x+\lambda y)(x+\lambda^2
y)\dots(x+\lambda^{n-1}y)\\=\sum\limits_{m=0}^n\left[{n\atop m}\right]_\lambda
\lambda^{\frac{m(m-1)}{2}}x^m y^{n-m}
\end{split}
\end{equation}
is determined by the coefficients
\begin{equation}
\left[{n\atop m}\right]_\lambda=\frac{[n]_\lambda!}{[m]_\lambda![n-m]_\lambda!}
 \label{9}
\end{equation}
where generalized factorials
$[n]_\lambda!=[1]_\lambda[2]_\lambda\dots[n]_\lambda$ are given by the basic
deformed numbers
\begin{equation}
[n]_\lambda=\frac{\lambda^n-1}{\lambda-1}.
 \label{10}
\end{equation}

\end{document}